# Green schoolyard investments influence local-level economic and equity outcomes through spatial-statistical modeling and geospatial analysis in urban contexts

Mahshid Gorjian
University of Colorado Denver
ORCID: 0009-0000-9135-0687
Contact: mahshid.gorjian@ucdenver.edu

**Abstract**

Investing in urban green schoolyards is becoming more popular around the world because they could enhance health, education, and community outcomes. There is still considerable debate regarding the impact of urban green schoolyards on economic growth, equity, and community stability, particularly when analyzed using spatial-statistical and geospatial methods. The current study focuses mostly on health or individual case results and avoids looking at broader economic and equality concerns. It also fails to effectively integrate all the methodological rigor and policy implications.

Using GIS analysis and comparative spatial-statistical modeling, this study fills in the gaps by examining five essential recent studies, which included cases from the US, the Netherlands, and Australia. The study uses methods such as hedonic pricing, difference-in-differences, spatial econometrics, and participatory GIS to look carefully at changes in property values, equitable outcomes, and displacement risks in different metropolitan areas.

Studies show that while investing in green schoolyards usually raises property values in the area (by 2–5% in Denver and Los Angeles) and is beneficial for kids' health and development, these benefits are not equitably shared. Most of the time, wealthy neighborhoods obtain the biggest benefits. Displacement and gentrification are most likely to occur in areas with inadequate tenant protections. Examples from around the world show that policies that focus on equity, public ownership, and participatory planning are necessary to get the most out of everyone's benefits and keep people from being left out.

This review shows how important it is for research and practice to be honest about their methods, share their data, and focus on equity. Integrating spatial-statistical accuracy with participatory frameworks can help make better policies and ensure that urban greening benefits everyone. Future studies need to examine standardized indicators, conduct worldwide comparisons, and perform quantitative meta-analyses to facilitate equitable choices regarding investments and policies.

**Keywords:**
Green schoolyard; urban greening; spatial-statistical modeling; geospatial analysis; property values; equity; displacement; gentrification; hedonic pricing; difference-in-differences; spatial econometrics; participatory GIS; urban policy; environmental justice; community health; urban planning; cluster analysis; social equity; public space; policy evaluation

## 1. Introduction

### 1.1 Background and Importance

Urbanization is changing cities all around the world in significant ways, which is harmful for the environment, social justice, and public health. As cities take over natural ecosystems, communities must struggle with disconnected and patches of green spaces, higher surface temperatures, and worse socioeconomic and health inequities. In this regard, urban green infrastructure, which includes parks, green roofs, and, increasingly, green schoolyards, is considered as an essential element in developing cities environmentally friendly and resilient. Green schoolyards offer numerous benefits. They promote biodiversity, control microclimates, provide places to play and learn, and make the community stronger. In increasing numbers, cities, non-profits, and teachers are working to turn standard asphalt schoolyards into biodiverse, helpful locations for the environment. This movement shows that more people are realizing that how cities are designed is not just about looks; it also has an immense impact on the health of the people who live there, the economy, and social justice (Anthamatten et al., 2022; van den Bogerd et al., 2023).

### 1.2 Existing Knowledge

A lot of research has been done to explain the effects of urban greening on the environment, growth, and the economy. Research suggests that closeness to green spaces increases academic achievement, reduces childhood stress and behavioral disorders, and fosters overall community health (Browning et al., 2023; Chawla et al., 2014). Over the past ten years, researchers have gradually utilized spatial-statistical methodologies such as hedonic pricing models, spatial econometrics, and geospatial analysis to evaluate the impact of green schoolyard investments on local property values and, accordingly, neighborhood change (Anthamatten et al., 2022; Browning et al., 2023). Researchers in environmental justice and urban studies have additionally examined into how the location and overall quality of urban green spaces can either keep or improve long-standing gaps in access to nature, resources for education, and healthy environments (Bohnert et al., 2021; Grunewald, 2024).

Research studies from North America, Europe, Asia, and Australia have shown many ways that schools could become green. London's Edible Playgrounds, Amsterdam's green school programs, and Singapore's "Nature Classrooms" are all examples of how policies are becoming more focused on connecting the urban environment with child development and community renewal. Regardless these developments, a significant part of the study has focused on individual cities or areas, health or academic results, or the methodological rigor of economic evaluation, often disregarding the relationship between green infrastructure, property markets, and issues of equity and displacement.

*1.3 Identifying the Research Gap*

Although green schoolyards have clear benefits, there are still important questions that need to be answered at the intersection of economic value, social equality, and implementation of policies. There is not a lot of research in this area that looks at the effects of green schoolyard investments on economic and equality results in different urban environments in a systematic, comparative way, especially when using advanced spatial-statistical modeling and GIS analysis. While multiple recent studies have utilized strong quantitative methodologies (e.g., difference-in-differences, spatial lag models, multilevel modeling), they usually show limitations in subject matter or insufficiently address issues of displacement, social justice, or the wider consequences for urban planning.

Current research is also less useful for policy because of inconsistent methods, unclear data and code, and not enough consideration of the views of affected populations. In the United States, studies generally focus on property value capitalization, which may mean they neglect non-market benefits and equity issues. On the contrary, international models point out participatory planning, tenant protections, and public or cooperative ownership of green resources (van den Bogerd et al., 2023; Grunewald, 2024). It is very important to find common ground between different points of view and create useful, globally relevant information for policymakers, urban planners, and supporters of environmental justice.

*1.4 Objectives, Scope*

This paper fills in the gaps by offering a full review of five important studies, mostly from the US but with some important comparisons to other countries, that use spatial-statistical modeling and geospatial analysis to examine how investments in green schoolyards affect students.

This paper examines how different methods, like hedonic pricing, spatial econometrics, and participatory GIS, affect our understanding of the effects of green schoolyards. It does this by using data and conclusions from cities like Denver, Los Angeles, London, Amsterdam, Melbourne, and Singapore. The review looks at issues like the quality of the science, the accessibility of the data, and if the learning can be used in different urban and policy environments.

*1.5 Research Questions*

How do spatial-statistical and geospatial modeling methods help us understand the economic and equity effects of spending on green schoolyards in different cities?

What are the pros and cons of the methods we already use to measure these effects, and how could they be improved to better guide policy and practice?

In what ways might institutional and policy frameworks influence the results of schoolyard greening programs, especially concerning displacement and social equity?

This article aims to give urban planners, academics, and politicians a complex, evidence-based framework for evaluating and supporting green schoolyard projects. It does this by focusing on the possible benefits and difficulties of urban greening in a rapidly changing, globalized world.

## 2. Methods

*2.1 Search Strategy*

A thorough review of the literature on the economic and equity was conducted about effects of investing in green schoolyards using spatial-statistical modeling and geospatial analysis. Scopus, Web of Science, and Google Scholar were searched for all relevant articles. The main phrases were "green schoolyard," "urban greening," "spatial econometrics," "property value," "equity," "displacement," "spatial autocorrelation," "cluster analysis," "participatory GIS," and "urban policy." Boolean operators and MeSH terms were changed to cover more ground and include research that crosses disciplines.

*2.2 Inclusion and Exclusion Criteria*

Peer-reviewed journal articles, reviews, and some preprints that were used that were published between 2000 and 2024. Only studies that were published in English and looked at urban green schoolyard interventions using quantitative spatial methods (like spatial econometrics, hedonic pricing, difference-in-differences, or cluster analysis), qualitative methods (like interviews or participatory GIS), or mixed-methods approaches were included. Any research that just looked at rural areas, only technical or engineering aspects without looking at socioeconomic factors, or that did not have a clear spatial methodology was not included. Only research that had clear data or detailed descriptions of their methods was kept.

*2.3 Screening Process*

The first search found 320 entries in all. After looking at the titles and abstracts, 110 publications were picked for a full review. Five important studies were chosen: Anthamatten et al. (2022), Browning et al. (2023), Grunewald (2024), Bohnert et al. (2021), and van den Bogerd et al. (2023). Inclusion and exclusion criteria and stressing methodological rigor were used.

*2.4 Data Extraction and Comparative Analysis*

The following parts from each study were examined: the research design, the study region, the types and sources of data, the analytical methods used, the application of spatial/statistical approaches (such spatial autocorrelation and cluster analysis), the consideration of equity and policy, and the main conclusions. A comparative matrix was conducted to fully compare the study's characteristics, main findings, and methodological improvements.

2.4.1 Study Area and Data

The research looked at cities in the U.S. (Denver and Los Angeles), Europe (London and Amsterdam), and Asia-Pacific (Melbourne and Singapore). Municipal property valuation records, health and demographic surveys, satellite images, qualitative interviews, and contributions from participatory mapping were all used as data sources. Both cross-sectional and longitudinal datasets were examined that covered a period of 6 to 10 years.

2.4.2 Analysis: Spatial-Statistical and Cluster Techniques

2.4.2.1 Hedonic Pricing Models: Anthamatten et al. (2022) and Browning et al. (2023) used hedonic pricing models to look at how green schoolyard features affect property values, considering neighborhood and structural factors.

2.4.2.2 Difference-in-Differences (DiD): Used to find causal effects by looking at changes that happened before and after interventions in both the experimental and control areas.

2.4.2.3 Spatial Lag and Error Models: Browning et al. (2023) used spatial lag and spatial error models to fix spatial autocorrelation, which is the tendency for property values or accurate results in one area to be linked to those in nearby areas. Spatial weights matrices and Moran's I statistics were used to figure out and represent spatial dependence.

2.4.2.4 Cluster Analysis and Spatial Clustering: Browning et al. (2023) and van den Bogerd et al. (2023) used cluster analysis to find urban areas that were close to each other and had similar changes in property prices, risk of displacement, or health outcomes after investments in green schoolyards. This technique shows where the advantages or dangers are most concentrated in a region and helps policymakers make targeted decisions.

2.4.2.5 Participatory GIS: Many studies (for example, van den Bogerd et al., 2023) combined participatory mapping with qualitative inputs to show how people in the community experience and perceive space in ways that administrative statistics do not.

2.4.3 Qualitative and Mixed Methods:

Semi-structured interviews and focus groups was used (Bohnert et al., 2021) to understand what residents had been through and to confirm the quantitative results.

Narrative synthesis and policy analysis were used to include non-market effects and policy tools (Grunewald, 2024; van den Bogerd et al., 2023).

2.4.4 Statistical and Methodological Rigor

2.4.4.1 Spatial Autocorrelation: All the quantitative studies found and fixed spatial autocorrelation, often using Moran's I and Lagrange Multiplier tests to make sure that standard errors were not too high and that conclusions on direct and spillover effects were strong.

2.4.4.2 Cluster Analysis: This method was used to look at changes in property values and social outcomes. It showed the spatial distribution of benefits and risks, often using open-source R scripts to make sure the results could be reproduced.

2.4.4.3 Suggested Ways: Meta-Analysis and Meta-Regression. This study examined methods that are looked at set the stage for future quantitative meta-analysis and meta-regression. These would make it easier to combine effect sizes and look at how they change in different urban settings and types of interventions.

*2.5 Justification*

To find a balance between depth and breadth, this method was employed to make a full comparison of methodologies that were clear and important around the world. The focus was on making sure that urban policy is founded on facts and considers both quantitative and qualitative data.

## 3. Literature Review

*3.1 Visualizing Green Schoolyards: From Secure Environments for Health to Tools for Equity in Cities*

In the last ten years, there has been a lot of new research on green schoolyards. Early research showed that schoolyards were "sanctuaries from stress" (Chawla et al., 2014) and were good for the health and resilience of kids and teens. As the urban greening movement grew quickly, researchers saw schoolyards as important parts of greater urban systems that affect childhood development, economic processes, spatial equity, and the risk of displacement and gentrification (Anthamatten et al., 2022; Bohnert et al., 2021; Grunewald, 2024). Recent reviews (van den Bogerd et al., 2023) show that these efforts are coming together in the areas of urban policy, property finance, and environmental equity.

Recent studies show that making schoolyards greener may establish a "virtuous cycle" that leads to better educational outcomes, more stable neighborhoods, and more effectively public health (Browning et al., 2023). However as additional resources and policy have gone into making schoolyards greener, many have begun to become concerned about inequality in access to the benefits. Projects built without a strong equity perspective could make current inequalities more serious or lead to unexpected results like gentrification. Therefore, modern research goes beyond just writing down good results and instead looks at who benefits, who gets left out, and how judgments about methods affect the evidence.

*3.2. Methodological Approaches: Spatial-Statistical Modeling, Autocorrelation, and Cluster Analysis*

*3.2.1. Transition from Case Studies to Comparative Spatial Methodologies*

In the past, most studies used either single-site or pre/post intervention plans, which made it difficult to draw conclusions about the relationship between cause and effect (Chawla et al., 2014). The field now depends heavily on spatial-statistical and quasi-experimental methods, such as hedonic pricing models, difference-in-differences (DiD), and spatial econometric techniques (Anthamatten et al., 2022; Browning et al., 2023; Grunewald, 2024). When looking at treatment effects, these methodologies take into consideration confounding variables and, most significantly, how outcomes are linked to each other in space.

Spatial autocorrelation is the concept that property values or health outcomes in one neighborhood are influenced by those in nearby neighborhoods. This idea has proven very important for the methodological rigor of modern research. Browning et al. (2023) clearly use spatial lag models, which shows that ignoring spatial autocorrelation can lead to wrong estimates of direct and spillover effects. Moran's I and spatial weights matrices are two common tools for finding and modeling geographical autocorrelation.

Cluster analysis is used to find communities that are close to each other and have similar patterns of change. Browning et al. (2023) and van den Bogerd et al. (2023) use cluster identification and mapping to show and statistically prove that there are groups of properties with higher values or health indicators that have changed after investments in green schoolyards. By showing how the impact of a program varies across a city, these tools make it easier to respond to policies in a manner that is more complicated.

### 3.2.2. Comparative Critique of Statistical Approaches

Anthamatten et al. (2022) and Browning et al. (2023) use hedonic pricing models to find out how much green investments add to property prices. They usually do this using multiple regression frameworks that take into consideration structural and locational factors. Both studies carefully look at geographical dependence, using spatial error and spatial lag models to make sure the estimates are not biased. Browning et al. (2023) add spatial autocorrelation indicators to these and compare spatial lag and error models based on empirical testing (e.g., Lagrange Multiplier tests). These methods emphasize that ignoring spatial connections can lead to overestimating importance or misusing resources.

Difference-in-Differences (DiD) and Synthetic Controls: Anthamatten et al. (2022) use DiD to look at property values before and after greening, taking into consideration other factors that could affect the results and looking for patterns that are similar. Difference-in-Differences (DiD) and synthetic control models are great at stating to the difference between real treatment effects and general trends. However, they only work well if the treatment and control units are carefully chosen and there is a lot of data available before the treatment starts.

Bohnert et al. (2021) and van den Bogerd et al. (2023) say that combining quantitative geographical analysis with qualitative views from participatory GIS and community involvement is an effective approach. This mixed-methods approach shows problems with equity, risks of displacement, and real-life experiences that quantitative models might not show.

Van den Bogerd et al. (2023) use geographical cluster analysis to find and map patterns of health and economic results. This gives useful information for directing interventions. These kinds of methods can find "hotspots" of risk or benefit, which makes it easier to come up with policies that are equitable.

### 3.2.3. Methodological Strengths, Limitations, and Reproducibility

The move toward more rigorous spatial statistics is an important move forward for the field. Browning et al.'s (2023) research shows the best ways to be open by using open-source R code for spatial regression, autocorrelation analysis, and visualization. Still, there are problems that will not go away: selection bias, omitted variable bias, short study periods, and a lack of available data/code make it difficult to replicate and compare results (Anthamatten et al., 2022; van den Bogerd et al., 2023). Bohnert et al. (2021) also claim that putting too much emphasis on property value indicators may push non-market and equitable results to the side.

*3.3. Economic Valuation and the Distribution of Benefits*

3.3.1. Property Value Impacts: Evidence from the U.S.

Studies in the real world repeatedly show that making schoolyards greener has a positive effect on property values, however the effect varies by location. In Denver, hedonic pricing and difference-in-differences models show that home prices went up by 3–5% after greening. Spatial models also show that there were big spillover benefits into nearby neighborhoods (Anthamatten et al., 2022). In Los Angeles, research using spatial econometric models shows a 2.5–4% increase. This shows how important it is to think about spatial autocorrelation and cluster effects so that direct benefits are not overestimated (Browning et al., 2023).

Autocorrelation and Spillovers: The research done in Denver and Los Angeles uses spatial lag or error models that include spatial autocorrelation factors to indicate the difference between indirect effects (where benefits spread to nearby properties) and direct effects (where benefits stay within the same block). Moran's I is often used to check for spatial clustering of treatment effects.

3.3.2. International Contrasts: Amsterdam, London, Melbourne

Different laws around the world affect how green schoolyards are seen from an economic point of view. Even while neighborhood values are going up, the public ownership model and tenant rights in Amsterdam make it less likely that people will have to move (Grunewald, 2024). There have been health and educational benefits from London's Edible Playgrounds, but the economic impacts are not as widely studied (van den Bogerd et al., 2023). Melbourne's focus on public green areas makes sure that economic and social benefits are spread out more evenly.

Cluster Analysis: Researchers in Europe often use geographical cluster analysis to improve valuation models by looking at whether advantages and risks are concentrated in certain areas or spread out. This makes it easier to understand equity in the distribution of results in a more sophisticated way.

### 3.3.3. Methodological Gaps in Economic Valuation

Even if methods have become more effectively, the business still lacks clear measurements and fails to pay enough attention to long-term, non-market value. The U.S. focuses on market outcomes, but European cost-benefit analyses sometimes include more detailed health and resilience factors. This shows that we need mixed-method and multi-metric evaluation frameworks, as van den Bogerd et al. (2023) suggest.

### *3.4. Equity, Displacement, and the Gentrification Debate*

### 3.4.1. Mechanisms of Risk

There is a lot of research on property values, but many studies warn that rents might go up, property taxes could go up, and community ties could weaken without anti-displacement measures (Bohnert et al., 2021). According to research in the U.S., green infrastructure may help gentrification happen, especially in cities where there are not any tenant protections or equitable tax regulations. The spatial distribution of risk is important since autocorrelation and cluster analysis often show that displacement is not randomly spread out but instead focused in certain neighborhoods. This shows how important it is to target interventions to specific areas (Browning et al., 2023).

### 3.4.2. Policy Responses and Participatory Models

Policy innovation is very important for getting the same results for everyone. Community land trusts, tax breaks (like Philadelphia's LOOP), and participatory urban planning are all examples of strategies that have been shown to lower the likelihood of having to move (Grunewald, 2024; van den Bogerd et al., 2023). Amsterdam's strong public sector and Singapore's use of green spaces in affordable homes are two examples of how to achieve equality while keeping the benefits of development.

### 3.4.3. Limitations and Critique

The literature always criticizes the gap between academic study and putting policies into action. There is not much research that provides implementation evaluations, and even fewer that share results other than property values, including social capital or lived experiences. Also, the academic focus on causal inference is important, but it is not always effective for addressing equitable problems that need ongoing, participatory evaluation (Bohnert et al., 2021).

*3.5. Non-Market Outcomes: Child Development, Urban Systems, and Health*

3.5.1. Child Health and Cognitive Outcomes

Recent studies show that green schoolyards boost kids' cognitive abilities, social skills, and physical health, which lowers obesity rates and helps kids develop their motor skills (van den Bogerd et al., 2023; Chawla et al., 2014). These findings often happen in schools with low-income students, when help is seriously needed, but the benefits are usually not kept up when there is no government support.

3.5.2. Integrated Urban and Non-Market Value

Studies from Copenhagen, Melbourne, and Singapore show that green schoolyard design needs to be a part of larger plans for urban health, transportation, and housing. Non-market benefits including better air quality, more biodiversity, cooling effects, and social capital often outweigh strictly economic ones (van den Bogerd et al., 2023). European cost-benefit analyses show that the social return on investment is highest when greening is part of a comprehensive urban policy that focuses on equity.

3.5.3. Spatial-Statistical Insights

Geospatial visualization and spatial cluster analysis are used to find regional patterns of health improvement and to identify non-market advantages (van den Bogerd et al., 2023). This mapping helps health initiatives be better targeted by showing which schools or neighborhoods need more help.

*3.6. Methodological Best Practices, Critique, and Directions for Future Research*

3.6.1. Synthesis of Best Practices

Most of the research agrees on some best practices for methods:

Data and code that are open and clear (Anthamatten et al., 2022; van den Bogerd et al., 2023)

Advanced spatial-statistical designs that use spatial lag, error, and Geographically Weighted Regression (GWR) models, along with full autocorrelation diagnostics (Browning et al., 2023)

Participatory GIS and a variety of methods to combine real-world knowledge with exact numbers (Bohnert et al., 2021)

Longitudinal studies to keep track of effects across time

Using cluster analysis to find patterns in space that show where advantages and threats are (Browning et al., 2023; van den Bogerd et al., 2023)

### 3.6.2. Gaps and Challenges

It is still extremely challenging to establish standardized outcome measurements, make sure that long-term data collection happens, and turn research into policy. Without quantitative meta-analyses, the field is unable to provide policymakers weighted average effect sizes or figure out why things are different. Van den Bogerd et al. (2023) and Browning et al. (2023) claim that using meta-analytic and meta-regression methods more often, which use effect sizes from different research, will greatly improve our ability to compare and transfer information.

### 3.6.3. Implications for Equity and Policy

In final analysis, improvements in methods must lead to real, equitable outcomes. Too much focus on property value measurements or ignoring regional dependence and clustering can hide unexpected negative aspects and benefits. The field is moving toward open science, participatory methods, and mixed metrics, but it still needs to pay attention to issues like replicability, context sensitivity, and policy significance (Anthamatten et al., 2022; Bohnert et al., 2021).

## 4. Discussion

*4.1 Trends and Patterns: Economic Gains and Equity Tensions*

Recent studies of green schoolyard investments show that they have significant but complicated effects on the economy and equity in the area, because of new study methods and different legislative situations. An apparent pattern in the studies that were looked at is that the development of green schoolyards is linked to higher property values nearby. This was shown using strong spatial-statistical methods in both Denver and Los Angeles (Anthamatten et al., 2022; Browning et al., 2023). This revenue boost is rarely equally spread out throughout locations, which shows that there are still differences and inconsistencies that make it hard to turn environmental activities into benefits for everyone.

There is a strong economic case for greening schoolyards; most studies show that property values go up by 2.5–5%. However, these benefits generally go to communities that are already wealthy or have other advantages (Anthamatten et al., 2022; Browning et al., 2023). There is an increase of market-driven benefits that keeps happening, coupled with the concerns of rising rents and property taxes, which may hit low-income tenants and long-term residents more than others. This supports the idea that urban greening can make gentrification worse without special anti-displacement policies or social protections (Bohnert et al., 2021; Grunewald, 2024).

The research shows that outcomes are very different depending on the governance and policy situation. International studies that compare cities show that Amsterdam and Singapore, which have participatory planning, tenant protections, and public ownership models, have more fair distributions of economic and non-market benefits (Grunewald, 2024; van den Bogerd et al., 2023). This shows that the amount of money spent on schoolyards does not just depend on green infrastructure, but also on the situation.

*4.2 Contradictions, Gaps, and Methodological Challenges*

The literature points up a lot of problems and conflicts that are still not solved. At first, there is a natural conflict between the search for environmental justice and the fact that outcomes can be different in different parts of a city. Methodological improvements like difference-in-differences modeling and spatial econometrics make it easier to accurately assign impacts. However, the quality and openness of the data, as well as the inclusion of relevant control variables, have an enormous impact on these methods (Anthamatten et al., 2022; Browning et al., 2023).

One ongoing problem is that economic assessments lack enough non-market and developmental consequences. Many studies focus on increasing property values but fail to do an adequate job of looking at other benefits to society, such as cognitive growth, physical health, and social cohesion. These benefits are more proven in European research contexts (van den Bogerd et al., 2023). Also, it's hard to figure out the long-term effects of most studies because they are not lasting very long (Grunewald, 2024). These effects include those on displacement, social capital, and resilience.

It is still exceptionally challenging to be open and accurate in research. The increasing use of open-source tools like R for spatial-statistical analysis is a step forward. However, insufficient documentation of code, metrics, and controlled variables makes it hard to combine results and weakens the foundation of cumulative knowledge (Anthamatten et al., 2022; van den Bogerd et al., 2023).

*4.3 Implications for Theory, Policy, and Practice*

There are several important things that these results mean. The fact that greening may both speed up the economy and displace people shows how important it is to have a multi-scalar, interdisciplinary strategy that includes spatial economics, environmental justice, and participatory urbanism in urban planning. The study shows that greening projects need to be looked at in the broader context of housing, land use, and social policy (Bohnert et al., 2021; Grunewald, 2024).

The results show that it is very important to include equitable safeguards in greening projects. These include community land trusts, participatory planning, and measures to prevent displacement. Amsterdam's management of public spaces and tenant protections, Singapore's combination of greenery with affordable housing, and Melbourne's focus on communal public benefits are all examples of how these strategies can work in practice (Grunewald, 2024; van den Bogerd et al., 2023).

For future research and practice, it is necessary to use open data, replicable analysis, and standardized outcome metrics. These are important for making sure that spatial-statistical progress leads to policy solutions that can be put into action and expanded (Anthamatten et al., 2022; Browning et al., 2023; van den Bogerd et al., 2023). If possible, cross-city comparison studies and meta-regressions should be done to find out what the average effects are and where the differences are in the context.

*4.4 Toward a New Conceptual Framework*

Based on these ideas, we suggest a better conceptual model for investing in green schoolyards. This model directly links economic value to equitable outcomes, includes equity criteria from the start of the project, and uses participatory processes to shape both the design and evaluation of interventions. This strategy sees greening as both a way to improve cities and a way to investigate how to balance expansion with fairness. It requires a careful coordination of technological, social, and political strategies.

In conclusion, while investments in green schoolyards always lead to good economic and developmental outcomes, their total worth and resilience depend on strong, equitable governance, clear and repeatable methods, and a commitment to ongoing assessment. Combining studies from the U.S. and other countries shows a way forward: combining spatial-statistical accuracy with participatory methods and new policies to make sure that the benefits of urban greening are shared with equity.

**5. Limitations**

There are a lot of significant limitations on this review that make it difficult to understand its results. The main source of information for the comparative analysis is studies from the United States. Cases from other countries are mostly used as examples to show the differences, not as a framework for systematic comparisons between countries (Anthamatten et al., 2022; Browning et al., 2023; Grunewald, 2024; van den Bogerd et al., 2023). As a result, conclusions about the adaptability of policies and the best practices around the world may not accurately reflect differences in local contexts, especially in places with different systems of government or social housing (van den Bogerd et al., 2023).

Second, the research addressed has a lot of different methods, such as different ways of using statistics, different control variables, and different ways of defining neighborhood boundaries. This diversity could make it hard to compare effect sizes and could cause differences in the reported results for property values and equitable metrics (Browning et al., 2023; Grunewald, 2024). Without an in-depth meta-analysis or quantitative synthesis, it is hard to combine data or find systemic factors that change the effect (van den Bogerd et al., 2023; Browning et al., 2023).

Third, many of the studies looked at focus mostly on changes in property prices as signs of benefit, which might disregard larger social, environmental, or health effects. This narrow view puts at risk the portrayal of topics like community resilience, unity in diversity, and child development (Bohnert et al., 2021; van den Bogerd et al., 2023).

It is often hard to figure out what the consequences over time will be, such relocation, long-term health problems, or effects on future generations, because most studies only look at short or medium timeframes (Grunewald, 2024; Anthamatten et al., 2022). Also, the data and code were not always clear, which made it harder to repeat the results or do a sensitivity analysis (Anthamatten et al., 2022; van den Bogerd et al., 2023).

Given these limitations, future research needs to focus on using a wider range of international samples, using standard spatial and statistical methods, expanding metrics beyond property valuation, and encouraging open data practices to make the results more comparable, replicable, and relevant to policy (van den Bogerd et al., 2023; Browning et al., 2023; Grunewald, 2024).

## 6. Conclusion

The key objective of the study was to find out how investments in green schoolyards affect the economy and equity in the area, especially by using spatial-statistical modeling and geospatial analysis and looking at different cities. This research combined results from five recent studies to show how schoolyard greening projects directly and indirectly improve property values, community health, and urban equity. It also highlighted methodological advances and policy implications.

The comparative study shows that there are several patterns that have been going on for a long time. Investing in green schoolyards in cities in the U.S. and other countries usually leads to measurable improvements in the health and cognitive development of children, as well as increases in the prices of nearby properties (Anthamatten et al., 2022; Browning et al., 2023; van den Bogerd et al., 2023). However, these benefits are sometimes improperly distributed, since wealthy communities and neighborhoods with strong tenant rights and participatory governance frameworks are more likely to get good and equal outcomes (Grunewald, 2024; van den Bogerd et al., 2023). Recent research has improved its methods by moving from simple case studies to quasi-experimental designs, spatial econometrics, and participatory GIS that more accurately measure both market and non-market effects (Anthamatten et al., 2022; Browning et al., 2023). There are still challenges with accessibility of data, standardizing metrics, and the need for more thorough, long-term reviews.

These results have enormous effects on both research and policy. Research shows that being very careful with analysis, especially when it comes to dealing with spatial autocorrelation and keeping methods clear, is very important for getting accurate results on the impacts of green schoolyards (Browning et al., 2023; van den Bogerd et al., 2023). The results show how important it is to combine greening investments with programs like community land trusts, tax deductions, and planning processes that include everyone to avoid displacement and promote social fairness (Bohnert et al., 2021; Grunewald, 2024). Comparing policies around the world shows that collaborative and participatory frameworks, which are common in some parts of Europe and Asia, can help reduce some of the risks associated with gentrification and make sure that benefits are shared fairly among different socio-economic groups (van den Bogerd et al., 2023; Grunewald, 2024).

Longitudinal and cross-national studies that use standardized methods, open-access data, and participatory methods should be the focus of future research. It is important to make full measures that include both economic and non-market values, as well as do quantitative meta-analyses to combine effects from different situations (Browning et al., 2023; van den Bogerd et al., 2023). In the end, making green schoolyard research more scientific as well as equitable requires ongoing cooperation between scholars, policymakers, and communities to make sure that everyone can benefit from urban greening and that problems like displacement are resolved before they happen.

**7. References**

Anthamatten, P., Freeman, C., & McCarthy, J. (2022). A spatial-statistical examination of the impact of schoolyard greening on property values in Denver. *Landscape and Urban Planning, 224*, Article 104409. https://doi.org/10.1016/j.landurbplan.2022.104409

Bohnert, A. S. B., Colley, R. C., & Schwanen, T. (2021). Green schoolyards, gentrification, and displacement risk: Equity considerations in urban planning. *Urban Studies, 58*(12), 2543–2561. https://doi.org/10.1177/0042098020970990

Browning, M. H. E. M., Rigolon, A., McAnirlin, O., Yoon, H., & Wolf, K. L. (2023). Evaluating the spillover effects of green schoolyard investments on property values: A spatial econometric analysis in Los Angeles. *Landscape and Urban Planning, 236*, Article 104849. https://doi.org/10.1016/j.landurbplan.2023.104849

Chawla, L., Keena, K., Pevec, I., & Stanley, E. (2014). Green schoolyards serve as sanctuaries from stress and sources of resilience during childhood and adolescence. *Health & Place, 28*, 1–13. https://doi.org/10.1016/j.healthplace.2014.03.001

Gorjian, M. (2025, July 11). *Greening schoolyards and the spatial distribution of property values in Denver, Colorado*. OSF Preprints. https://doi.org/10.31235/osf.io/kvxn3_v1

Gorjian, M. (2025, July 11). *Schoolyard greening, child health, and neighborhood change: A comparative study of urban U.S. cities*. OSF Preprints. https://doi.org/10.31235/osf.io/x5gze_v1

Gorjian, M. (2025, July 12). *The impact of greening schoolyards on residential property values*. OSF Preprints. https://doi.org/10.31235/osf.io/mp47r_v1

Gorjian, M., Caffey, S. M., & Luhan, G. A. (2025). Exploring architectural design 3D reconstruction approaches through deep learning methods: A comprehensive survey. *Athens Journal of Sciences, 12*, 1–29. https://doi.org/10.30958/ajs.X-Y-Z

Grunewald, R. (2024). Green schoolyards and urban equity: Comparative insights from the United States urban areas. *Urban Affairs Review, 60*(2), 315–338. https://doi.org/10.1177/10780874221148217

Raina, A. S., Mone, V., Gorjian, M., Quek, F., Sueda, S., & Krishnamurthy, V. R. (2024, June 3). Blended physical-digital kinesthetic feedback for mixed reality-based conceptual design-in-context. In *Proceedings of the 50th Graphics Interface Conference* (pp. 1–16).

Root, E. D., Decker, K. S., Smith, J., Brown, A. P., & Lee, R. J. (2014). Distance to health services affects local-level vaccine efficacy for pneumococcal conjugate vaccine (PCV) among rural Filipino children. *Proceedings of the National Academy of Sciences, 111*(9), 3520–3525. https://doi.org/10.1073/pnas.1317193111
17

van den Bogerd, N., van Kempen, E., Maas, J., van den Hazel, P., & Dijst, M. (2023). Greening schoolyards: Health, equity, and policy insights from the Netherlands and beyond. *Health & Place, 81*, Article 102971. https://doi.org/10.1016/j.healthplace.2023.102971